\documentclass[12pt]{emulateapj}
\usepackage{graphicx}
\usepackage{epsfig}
\usepackage{subfigure}
\usepackage{amsmath}    
\usepackage{longtable}
\usepackage{dcolumn}
\usepackage{amssymb}
\usepackage{natbib}

\usepackage{rotating}

\slugcomment{Submitted to ApJ, January 31, 2005}

\shorttitle{The Milky Way Spiral Arm Dynamics }
\shortauthors{Naoz \& Shaviv}

\begin{document}

\def\omunit{\mathrm{km~sec^{-1}~kpc^{-1}}}
\def\omunits{{\small km~sec$^{-1}$~kpc$^{-1}$}}

\def\vunit{\mathrm{km~sec^{-1}}}

\def\mvec{\mathbf}
\def\mr{\mathrm}

\def\figsize{4.0in}
\def\figsizesmall{3.0in}


\title{ Open Cluster Birth Analysis and \\  Multiple Spiral Arm Sets in the Milky Way  }


\author{Smadar Naoz and Nir J.\ Shaviv}
\affil{Racah Institute of Physics, The Hebrew University, Jerusalem 91904, Israel}







\begin{abstract}



The Milky Way spiral arm dynamics is studied using the birth place of open clusters. We separately study the nearby spiral arms, and find evidence for multiple spiral sets. In particular, the Sagittarius-Carina arm appears to be a superposition of two sets. The first has a pattern speed of $\Omega_{P,\mr{Carina},1}=16.5^{+1.2}_{-1.4sys}\pm1.1_{stat}~\omunit$, while the second has $\Omega_{P,\mr{Carina},2}=29.8^{+0.6}_{-1.3sys}\pm1.3_{stat}~\omunit$. The Perseus arm located at larger galactic radii exhibits only one clear set with $\Omega_{P,\mr{Perseus}}=20.0 ^{+1.7}_{-1.2sys}\pm1.6_{stat}~\omunit$.
The Perseus and ``slower" Carina arms are most probably part of an $m=4$ set. The Orion ``armlet" appears to have a pattern speed of $\Omega_{P,\mr{Orion}}=28.9 ^{+1.3}_{-1.2sys}\pm0.8_{stat}~\omunit$. In other words, both the ``faster'' Carina arm and the Orion arm are nearly corotating with the solar system.

\end{abstract}


\section{Introduction}
\label{sec:intro}

There is currently no doubt that the Milky Way is a spiral armed galaxy. Yet,  our edge-on view and dust obscuration through the galactic disk do not permit us to see the structure and pattern
 of the spiral arms  directly. Thus, different claims for different spiral arm geometries and dynamics are abundant in the literature. Claims for a double or 4-armed structure are common, while numerous estimates for the pattern speeds can be found.   
 
A 4-armed spiral pattern does however appear to be preferred by the observational data \citep{Georgelin,Vallee}. In particular, velocity-longitude maps reveal quite clearly the existence of a 4-armed spiral pattern {\em outside} the solar circle \citep{Blitz,Dame}\footnote{The maps reveal 3 arms separated by $\sim 90^\circ$, a fourth arm is presumably obscured behind the center of the galaxy.}. Unfortunately, however, nothing equivocal can be stated at smaller galactic radii because of the inherent ambiguity in $v-\ell$ maps and the large scatter in the indirect tracers of spiral arms.  

Besides the number of arms $m$, in the spiral set, the spiral arms are also characterized by a pitch angle $i$, a pattern speed $\Omega_P$ and an amplitude. For a  logarithmic arm, the pitch angle is fixed and given by
$
\cot i=\left|{R ({\partial\phi}/{\partial R})}\right|,
$  
where $R(\phi)$ is the parametric description of the arm. By considering the various observations at the time, \cite{Vallee} has shown that an $m=4$ set, with a pitch angle of $i = 12 \pm 1^\circ$, is notably more consistent with the data.
  

Notably less consensus, however, can be found for the value of the pattern speed of the spiral arms. Estimated values, based on different methods and data sets,  typically range between 10 to 30 $\omunit$  \citep[for a recent review, see][]{ice}, but the distribution is not even---the $\sim 20$ results  seem to divide  into three different  ranges of values. These are  $\Omega_\odot-\Omega_p \sim 9-13, \sim 1-5$ and $\sim (-1) - (-4) ~\omunit$.  

We can classify the methods according to the core assumptions upon which they are based.  In particular, several methods implicitly assume that the spiral arms are necessarily a density wave \citep{LinShu1,galacticD}, and a few even assume a particular origin of the spiral arm perturbation.  

For example, \cite{lys} argued that the only solution possible for spiral arms are those  described by the aforementioned density wave theory. Under small perturbations, a best fit  $\Omega_P=11-13~\omunit$ was obtained. \cite{yuan2}  found that the observed 21-cm line profile is consistent with the $\Omega_P$ range found by \cite{lys}.  However, \cite{nelson} compared the shock solution for the profile of the gas perturbed by a spiral wave to the 21-cm line observed in the  direction opposite to the center of the galaxy,  and derived that $\Omega_P\approx20~\omunit$. The explanation for the different results, according to \cite{nelson}, is that \cite{yuan2} used a linear theory for the gas motion, and that the \cite{lys} results could have been modified by the existence of a high velocity halo.

\cite{second} modeled the Milky Way as a barred galaxy and allowed separate pattern speeds for the bar and for the spiral structure. By using hydrodynamic simulations and comparison to the observed $v-\ell$ maps in CO, they showed that the spiral arms penetrate the bar's corotation region, such that within the same region of the bar, there are two different pattern speeds.
The bar's pattern speed obtained is $\approx59  \pm5~\omunit$, and that of the spiral arms  is $\approx19\pm5 ~\omunit$ (while assuming that the sun is located at $8$~kpc from the galactic center). 
  
A second method which still relies on the assumption that the spiral arms are described by the density wave theory, is through the study of the {\em kinematics} of various populations, which orbit the galaxy in orbits modified by the spiral wave.  Because the method relies only on kinematics, the determination of the age of the spiral tracers is not important as it is for other methods (e.g., when using star cluster gradients as is done here). \cite{comeron} obtained $\Delta\Omega_P \equiv \Omega_\odot - \Omega_P =10\pm5~\omunit$, thus $\Omega_P =16\pm5~\omunit$, and a pitch angle of $i\approx 8^{\circ}$ for a sun located at $8.5$~kpc from the galactic center, with an angular speed of $220~${km}~{sec}$^{-1}$.

Another method which still relies on the density wave theory, is the identification of resonance features.  \cite{gordon}  determined the existence of a discontinuity in the CO abundance as a function of  radius, at the region of $R=4~kpc$, and identified it with the inner Lindblad resonance. This result is consistent with  $\Omega_P=11.5\pm1.5~\omunit$, for a two spiral-armed Milky Way.  Gordon's analysis was for the region of $2~\mathrm{kpc} < R < 10~\mathrm{kpc}$.
 
In a second type of methods,  the physical origin of the spiral arms is not assumed. Instead, objects which are linked to spiral arms and can be ``dated'', are analyzed. Specifically, either the age gradient or the birth place of particular objects can be studied to reconstruct the dynamics of the spiral arms. For example, \cite{palous} fitted the place of birth of open clusters to a modeled Milky Way based on 21-cm observations. They obtained  two possible values: $\Omega_P\approx13.5~\omunit$ and $\Omega_P\approx20~\omunit$. 

Interestingly, various authors actually found multi-pattern speeds co-existing in the galaxy. \cite{multi1} presented a two dimensional model for multiple pattern speeds in barred galaxies. Their results  are in agreement with \cite{multi2} and  \cite{multi3} for barred galaxies. \cite{bar6}, \cite{bar5} and many other observations  support the possibility of bar in the Milky way, which may explain the origin of multiple harmonics in the arms. 

To summarize, the current determinations of $\Omega_P$ are inconclusive. In this paper, we will try to estimate the value of $\Omega_P $ using the birth place method, for young clusters (age $\leq 5 \times 10^{7}$~yr), without limiting ourselves to assumptions of a particular theory, such as the density wave theory. We will show that the origin of the large confusion rests in the existence of at least two spiral sets besides the Orion armlet, within which the solar system currently resides. 

The last parameter characterizing the spiral arms is their amplitude. We will not dwell on this point here, but a more elaborate analysis aimed at finding the finite amplitude based on the same set of open clusters, is underway.

\section{Spiral Structure Analysis  }\label{ssa}
Here, we present the  methods used  to map and study the dynamics of the Milky Way spiral arms in our galactic vicinity. In our analysis, we used the \cite{dias} and \cite{loktin4} databases of open clusters. These include the galactic coordinates and age estimates for open clusters at typical distances of up to a few kpc from the sun. We used \cite{olling} for the rotation curves. 

We work in cylindrical galactic-centered coordinates, $r$, $\phi$ and $z$. The solar system's position in the Milky Way has not been accurately ascertained yet, conservatively, it ranges between $7.1$~kpc and $8.5$~kpc, based on velocity-longitude maps of CO data \citep{olling}. Nevertheless, recent determinations based on the dynamics of stars around the central black hole do give a tighter estimate. For example, \cite{Eisenhauer} find a distance of $r_\odot = 7.94 \pm 0.42$~kpc based on the observed dynamics of stars around the central black hole, while using Hipparcos data, \cite{SG} find $r_\odot = 8.2 \pm 0.15_{stat}\pm 0.15_{sys}$~kpc. We will be conservative henceforth and consider that the larger range of $7.1-8.5$~kpc is possible, but keep in mind that the higher values are more likely.  

Similarly, the reasonable solar rotational velocities $v_\odot$ range between $184$ and $200$\ $\vunit$ for a solar system at $r_\odot = 7.1$~kpc, and $220 - 240~\vunit$ for $r_\odot = 8.5$~kpc. We carry out 
all our analyses while assuming five possible locations and velocities. The 
parameters of these five different possible configurations for the sun orbiting the galaxy are $r_\odot=7.1$~kpc and
 $r_\odot=8.5$~kpc with the lower and upper $v_\odot$ of the two aforementioned ranges,
  and $r_\odot=7.8$~kpc with $v_\odot=202~\vunit$ as a nominal case.

We conduct our analysis in a frame of reference rotating with the spiral pattern speed, $\Omega_P$. In this frame, the differential angular velocity of each cluster is given by
 \begin{equation}
\Delta\Omega=\Omega(r)-\Omega_P.
 \label{delta}
\end{equation}

Furthermore, we consider only relatively young clusters, having ages satisfying $t \leq 5 \times 10^{7}$~yr. This implies that we preform our analysis only on clusters which did not have time to move too far away from the arm\footnote{Typically for clusters, $\Delta\Omega\lesssim13~\omunit$ (see \S\ref{sec:carina}), therefore, the time period for a rotation in the spiral arm frame of reference is $\gtrsim 6 \times 10^{8}~$yr. For a four arm set, clusters will cross arms every $\gtrsim 1.5 \times 10^{8}~$yr. Hence, in $\lesssim 5  \times 10^{7}$~yr, typical clusters  will not be  displaced much from the arm.}.



The most detailed $v-\ell$ maps of the Milky Way are the \cite{Dame} observations of molecular gas using CO as a tracer. In this data set, one can clearly identify three different arms: The Carina arm which is located inwards to the solar galactic radius $r_{\odot}$, the Perseus arm, externally located relative to $r_\odot$, and the ``outer arm" beyond the Perseus one. Our Sun is located in the Orion arm, which is probably only an ``armlet" \citep{Georgelin}.

We assume the most general case in which the observed arms are not necessarily part of the same set of spiral arms. As we shall demonstrate, this is necessary considering the multiple pattern speeds which appear to be present.
Therefore, we analyze each arm separately, and choose to work only with young star clusters (having $t \leq 5 \times 10^7$~yr, as explained above). We perform our analysis on the \emph{Carina} arm, the \emph{Orion} armlet and the \emph{Perseus} arm. 

We work in the frame of reference of the arms. In this frame of reference, the birth location is given by\footnote{This neglects the finite amplitude of the spiral density wave. In our ongoing research, we alleviate this assumption by including the epicyclic motion arising from the finite spiral arm perturbation. Also, as young objects, the peculiar motion of the open clusters is negligible.}:
\begin{equation} 
\phi_{birth}=\phi-[\Omega(r)-\Omega_P]\cdot t,
\label{birth}  
\end{equation}
where $t$ is the age of the clusters, and $\Omega(r)$ was taken from \cite{olling}. We find a best fit for three parameters: $\Omega_P$, the inclination angle $\tan(i)=\big|{d(\ln r)}/{d\phi}\big|$ and the intersection point with the $\phi$-axis. For each $\Omega_P$ we find the best fit, according to a weight function defined below.

\subsection {Statistical Methods}
We continue now with the description of the algorithm we employ, and in particular, the statistical methods used. We begin by assuming a logarithmic arm profile \citep{RRS}, given by 
\begin{equation} 
r=e^{a\phi_\mr{birth}+b}.
\label{log_arm}  
\end{equation}
Here $a$ is related to the pitch angle $i$ through the relation  $i=\tan^{-1}(a)$. In other words, $a$ describes the slope of the spiral arm in the $\ln{r}$ vs.\ $\phi_{birth}$ plot, while $b$ is the intersection point with the $\ln{r}$ axis. 

We wish to find the best fit for a logarithmic arm. Thus, we define a weight function ${\cal W}$ with which we evaluate the goodness of the fit. The definition we chose is
\begin{subequations}
\label{w}  
\begin{eqnarray}
{\cal W}(a,b)=-\sum_{i=1}^{N} \max\left( \big[1-\frac{\Delta_i}{d}\big],0\right) \label{wa}\\
 \Delta_i= \big| \ln r_i-(a\phi_{birth,i}+b)\big|, 
\label{wb}
\end{eqnarray}
\end{subequations}
where $N$ is the number of data (i.e., cluster)  points, and $d$ is a scaling factor describing the typical width of the arm. The argument in the summation is positive for a cluster inside the arm, and larger for a cluster closer to the center. The condition $\Delta_i\leq d = 0.05$ enables us to disregard data points which do not lie within $\sim 700$~pc. In other words, we effectively consider the width of the arms to be $700$~pc. The advantage of such a choice for ${\cal W}$ over a more standard definition (such as least squares), is that under a standard definition such as the sum of distances squared,  outliers are very important for the minimization. However, in our case, it is most likely that outliers are simply unrelated to the arm and should therefore be disregarded altogether. 
This is because the definition is such that when $\ln( r_i)=a\phi_{birth ,i}+b$, the argument in the summation obtains its maximum value of 1,  while for $ \left|\ln( r_i)-(a\phi_{birth ,i}+b)\right| \geq  d$ the argument obtains its lowest value 0, such that the cluster is disregarded. 

The weight function ${\cal W}$ is defined using the distances in the $\ln r $ axis, since in any usage of $\phi_\mr{birth}$ (as defined in eq.~\ref{birth}) involves the age of the cluster which has a higher error. 
The best fit for a given $\Omega_P$ is then obtained by minimizing ${\cal W}$. A best fit is subsequently obtained for different $\Omega_P$ as well. 

We considered two minimization methods for ${\cal W}$; these were the metropolis simulated annealing method \citep[e.g.][]{NU}  and a ``brute force" method, where we loop over the whole parameter space with a finite resolution. The advantages of the former method are the smaller CPU requirements and higher accuracy reached. The disadvantage, on the other hand,  is the uncertainty of whether an obtained minimum is indeed the best fit. For this reason, we also performed a ``brute force" minimization by looping over parameter space with a finite resolution, to ensure that the simulated annealing reached the absolute minimum. 

The brute force method also proved useful because the additional local minima could be identified. As we shall see below, an additional minimum is related to an additional set of arms, coexisting in the data.

Since we have no or very poor knowledge of the statistical properties of the weight function (eq.~\ref{w}), nor of the statistical or physical variance in the cluster data, we used the {\em Bootstrap Method} \citep[e.g.,][]{NU} to estimate the confidence levels around the obtained minima. This method allows us to  evaluate the errors of the obtained results using the data itself, hence its name.  

Specifically, we estimate the error in the best fit as follows. Our best estimate for the best fit parameters is obtained by minimizing the weight function ${\cal W}$ using the original cluster data. To estimate the error on the best fit parameters, 
we then build a set of different realizations of the data, which differ from the original set by having  a random $37\%\sim {1}/{e}$  of the clusters replaced with random clusters chosen from the original set. We then apply the minimization procedure to each new data set. The estimated best fit parameters in each realization are expected to vary from the best estimate (based on the original set) by $\sim 1 \sigma$. Thus, we can use the ensemble of modified sets to calculate the errors.   

\def\figureone{
\begin{figure}[t]
\centering
\epsfig{file=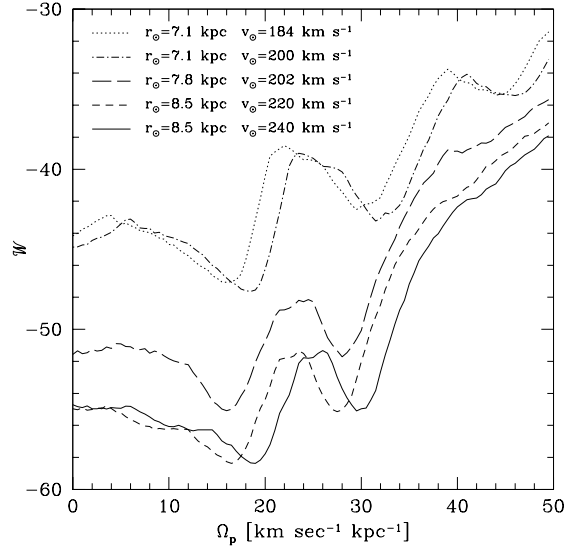, height=\figsizesmall}
\caption{\footnotesize Weight function vs.\ the pattern speed, $\Omega_P$, for the  lower (Carina arm). The five graphs represent different assumed $r_\odot$ and $v_\odot$. Each graph is obtained by minimizing the weight function while fixing the value of $\Omega_P$. The two minima in ${\cal W}$ demonstrate that there are probably two sets of spiral arms.}
\label{fig:h_lower}
\end{figure}
}
\figureone

\section{The Dynamics of the Carina Arm}

\label{sec:carina}
 
We now proceed to apply the above methods to the Carina arm. We begin by calculating the best fit for the spiral arms as a function of the assumed $\Omega_P$.  We plot the minimized weight function 
${\cal W}$ as function of $\Omega_P$ and find a bimodal pattern (see fig.~\ref{fig:h_lower}). This can be interpreted as \emph{two different sets of spiral arms} in the Milky Way.


A second apparent feature of fig.~\ref {fig:h_lower} is that minimization fits having a larger solar-galactocentric radius yield overall lower minima. This strongly suggests\footnote{This assumes that the spiral arms are logarithmic spirals, as favored in the density wave theory \citep{RRS}. If this assumption is alleviated, lower galactocentric radii could still be possible.} that $r_\odot \gtrsim 8$~kpc. Interestingly, it is consistent with various recent determination of the distance to the galactic center, as mentioned in \S\ref{sec:intro}.
 
\noindent{\bf The First Set:}
From the two minima we find (see fig.~\ref{fig:h_lower}), the minimum corresponding to the lower pattern speed in the angular velocity ``spectrum" appears to be a somewhat better fit. The results for the different fits to a logaithmic arm, while assuming different solar galactocentric radii and velocities, are depicted in fig.~\ref{fig:lower_firstarm}. The actual values of this best fit (henceforward, the ``slower Carina arm") are summarized in table \ref{tab:all}.

\noindent{\bf The Second Set:}
The Carina-arm cluster data include a second statistically significant minimum. The significance can be inferred from the fact that all realization which have $1/e$ of the clusters randomly exchanged with clusters from the data set still exhibit this minimum, albeit with a slightly modified best-fit parameters. We therefore interpret the data as two existing spiral arms with two different velocities which happen to spatially overlap at present. The actual best-fits for this solution (henceforward, the ``faster Carina arm") are plotted in  fig.~\ref{fig:lower_secondarm}, and summarized in table~\ref{tab:all}. 

Note that although the pattern speed here is faster than in the first set, it is close to the orbital angular velocity of the sun, implying that the ``synodic" angular velocity of the faster set is relatively small.

As summarized in \S\ref{sec:intro} and elaborated in \cite{ice}, the determinations of $\Omega_p$ appear to divide  into three different ranges of pattern speeds. The two values obtained for  $\Omega_P$ in the Carina arm (see fig.~\ref{fig:h_lower}) nicely agree with two of the three different ranges obtained by other authors. We will elaborate on the implications of the results in the discussion.

\def\figuretwo{
\begin{figure*}
\centering
\plottwo{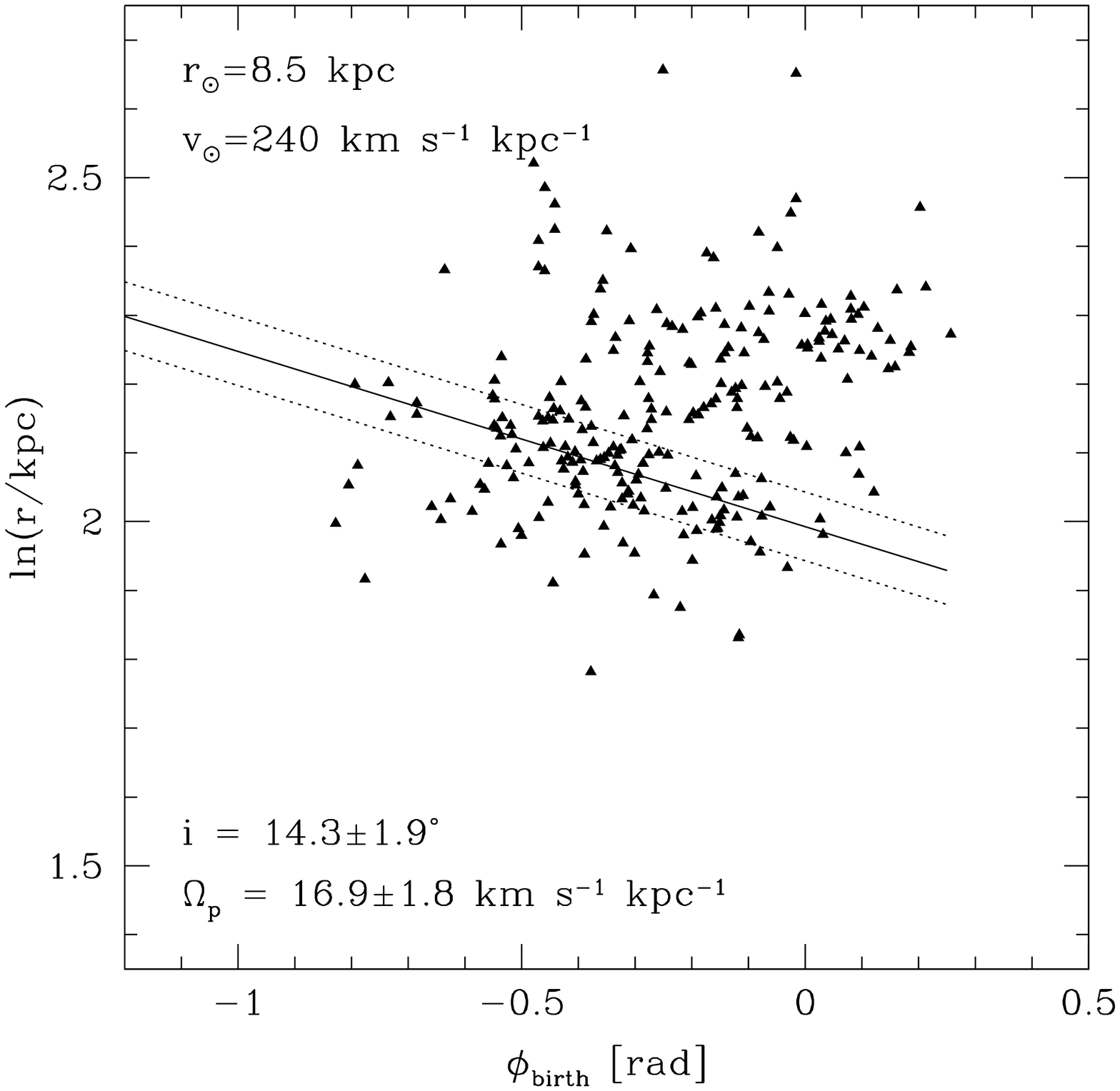}{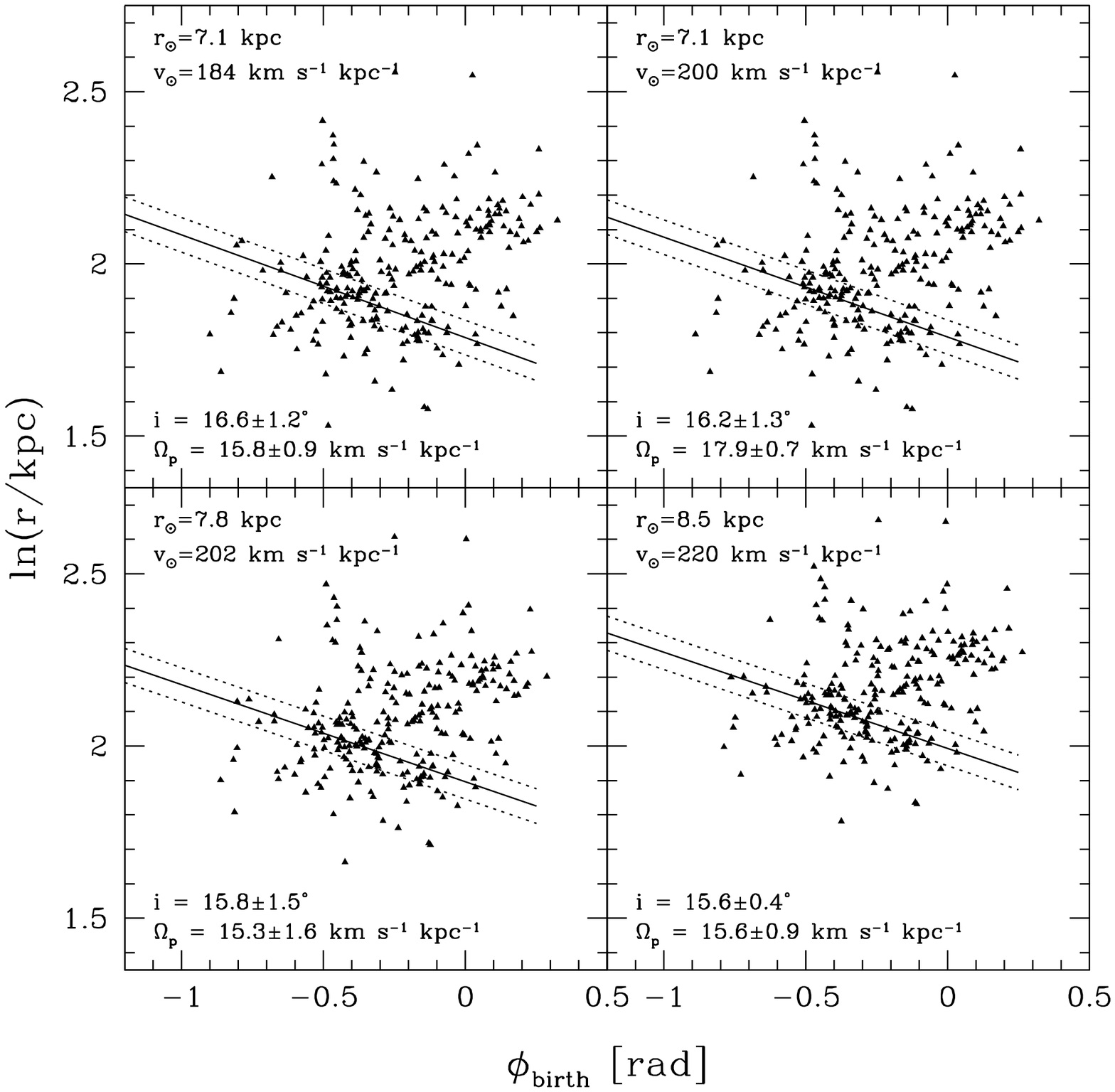}
\caption{\footnotesize A plot of $\ln r$ vs.\ the birth angle $\phi_\mathrm{birth}$ of the open clusters and the best fit for the first Carina arm, assuming different $v_\odot$ and $r_\odot$. The solid line depicts the best fitting logarithmic arm, while the dotted lines denote $\pm 700$~pc strips delineating the clusters which are considered as part of the spiral arm.}
\label{fig:lower_firstarm}
\end{figure*}
}
\figuretwo

\def\figurethree{
\begin{figure*}
\centering
\plottwo{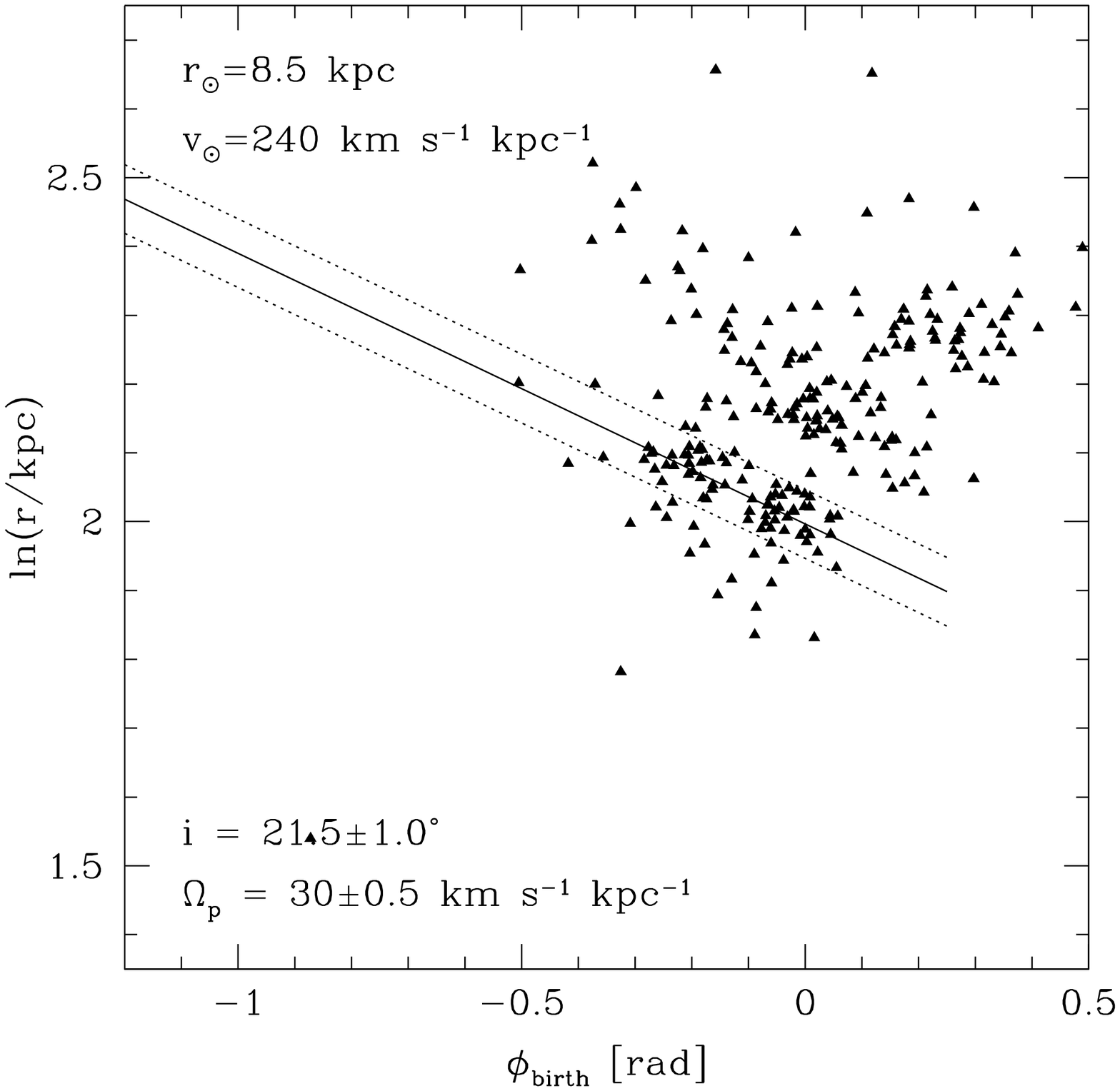}{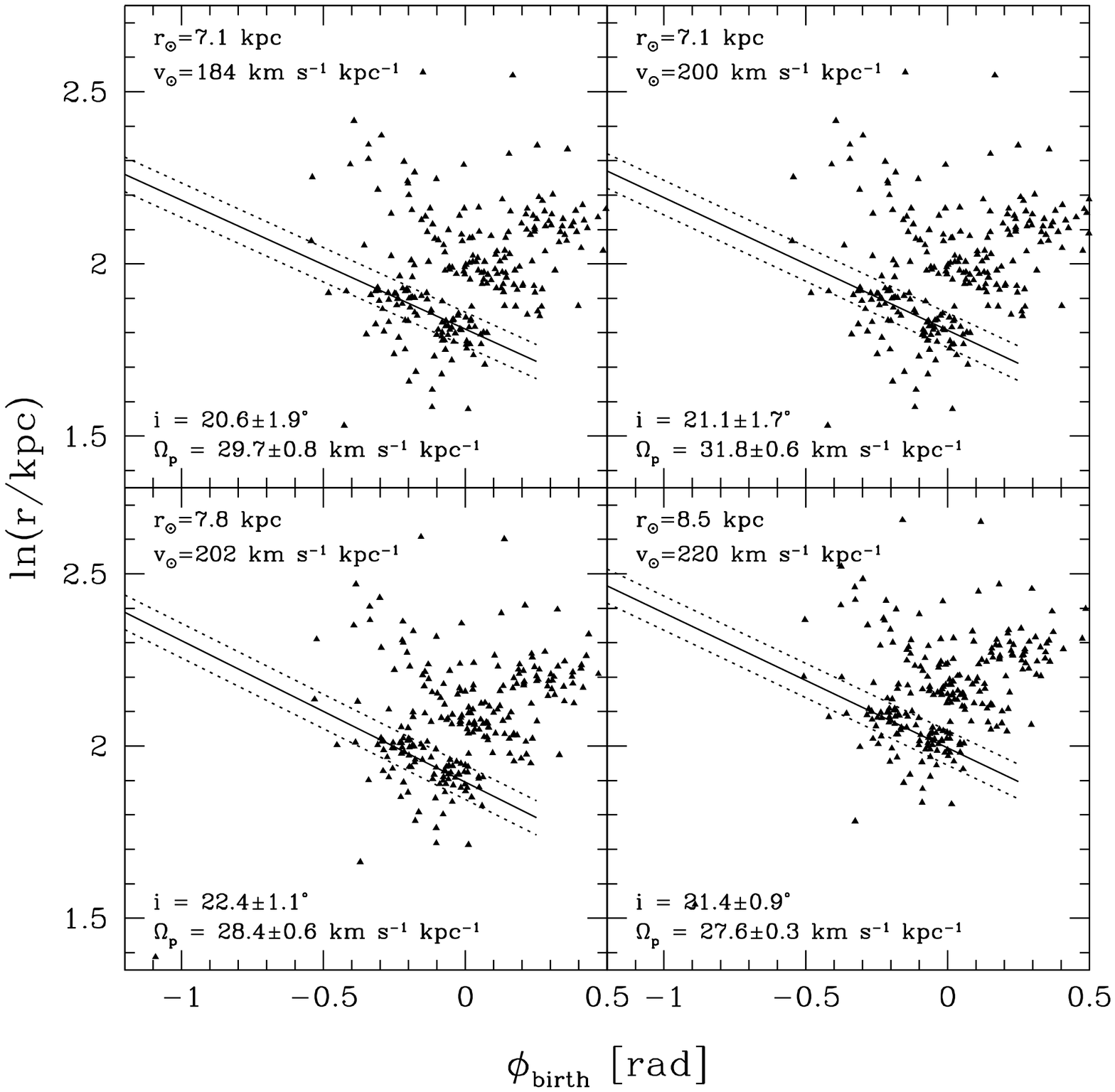}
\caption{A plot of $\ln r$ vs.\ the birth angle $\phi_\mathrm{birth}$ of clusters and the best fit for the second Carina arm, similar to fig.~\ref{fig:lower_firstarm}.}
\label{fig:lower_secondarm}
\end{figure*}
}
 \figurethree

\def\figurefour{
\begin{figure}[t]
\centering
\plotone{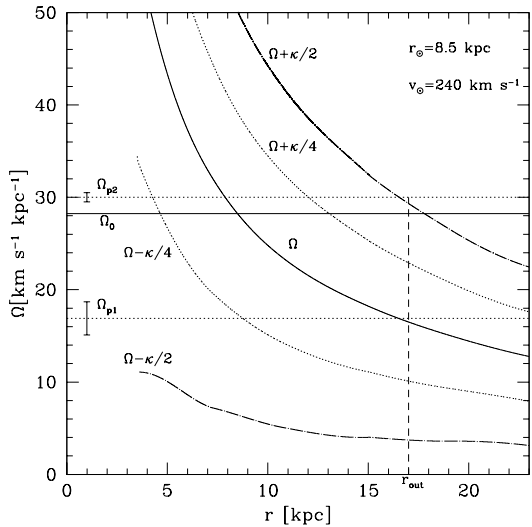}
\caption{\footnotesize The circular frequency $\Omega$ (solid line) and the 1:2 (dash-dotted line) and 1:4 (dotted line) Lindblad resonance frequencies as a function of the galactic radius. Given that spiral arms are observed  out to $r_\mr{outer}\approx 2r_{\odot}$ \citep{Blitz}, this is probably the outer Lindblad resonance. Nevertheless, the outer Lindblad resonance could in principle also be at larger radii. The four arms could be part of one set with $\Omega_{P1}$. In this case, the outer extent of the spiral arms, should be identified not with the outer Lindblad resonance, but with the co-rotation. Interestingly, here the inner Lindblad resonance is near $r_\odot$. We also see that as most two of the arms could be part of an $m=2$ set with $\Omega_{P2}$. In this case, the outer extent of the spiral arms is just near the expect outer Lindblad resonance.   }
\label{fig:resonances}
\end{figure}
}
\figurefour

\def\figurefive{
\begin{figure*}[t]
\centering
\plottwo{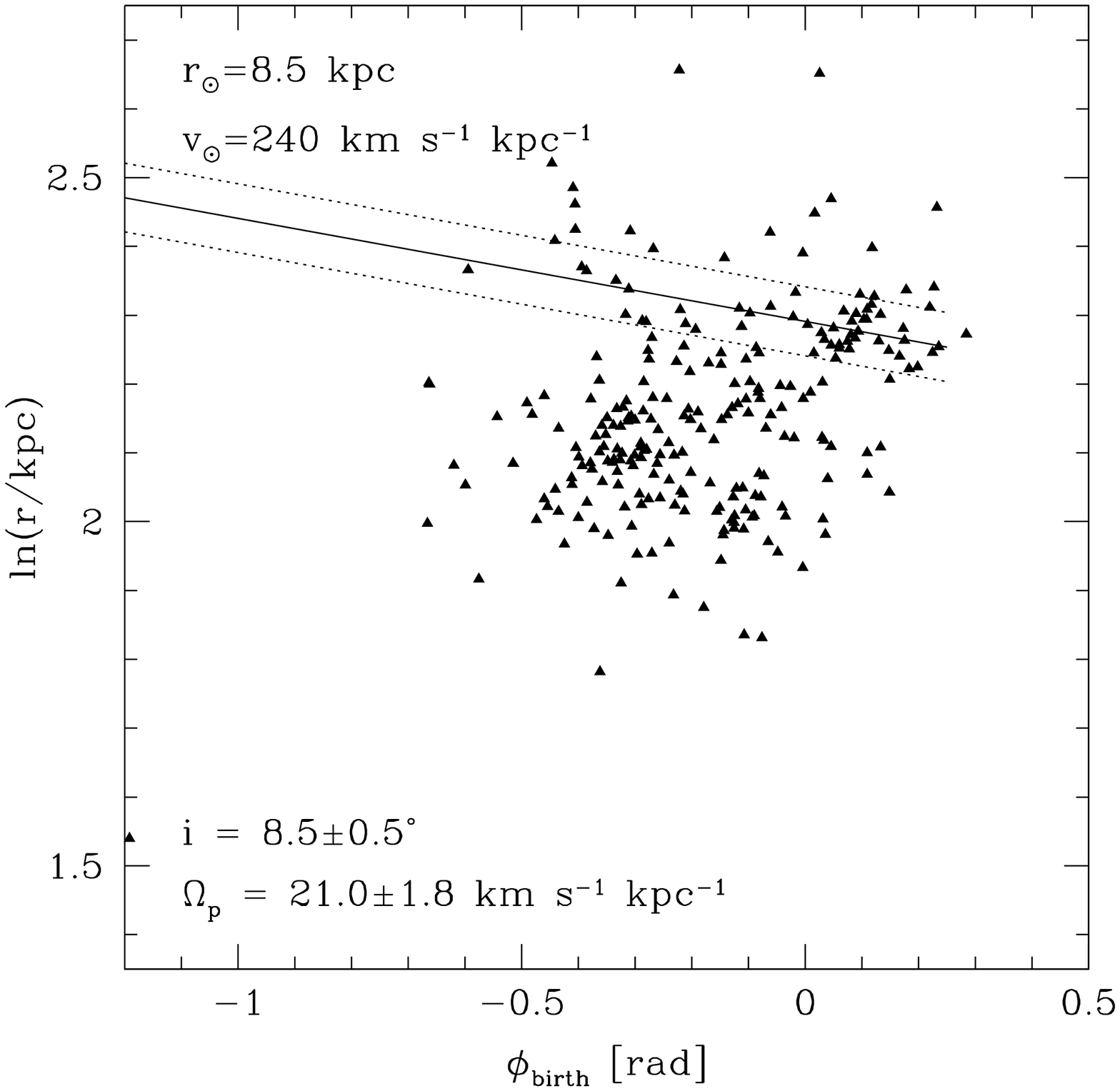}{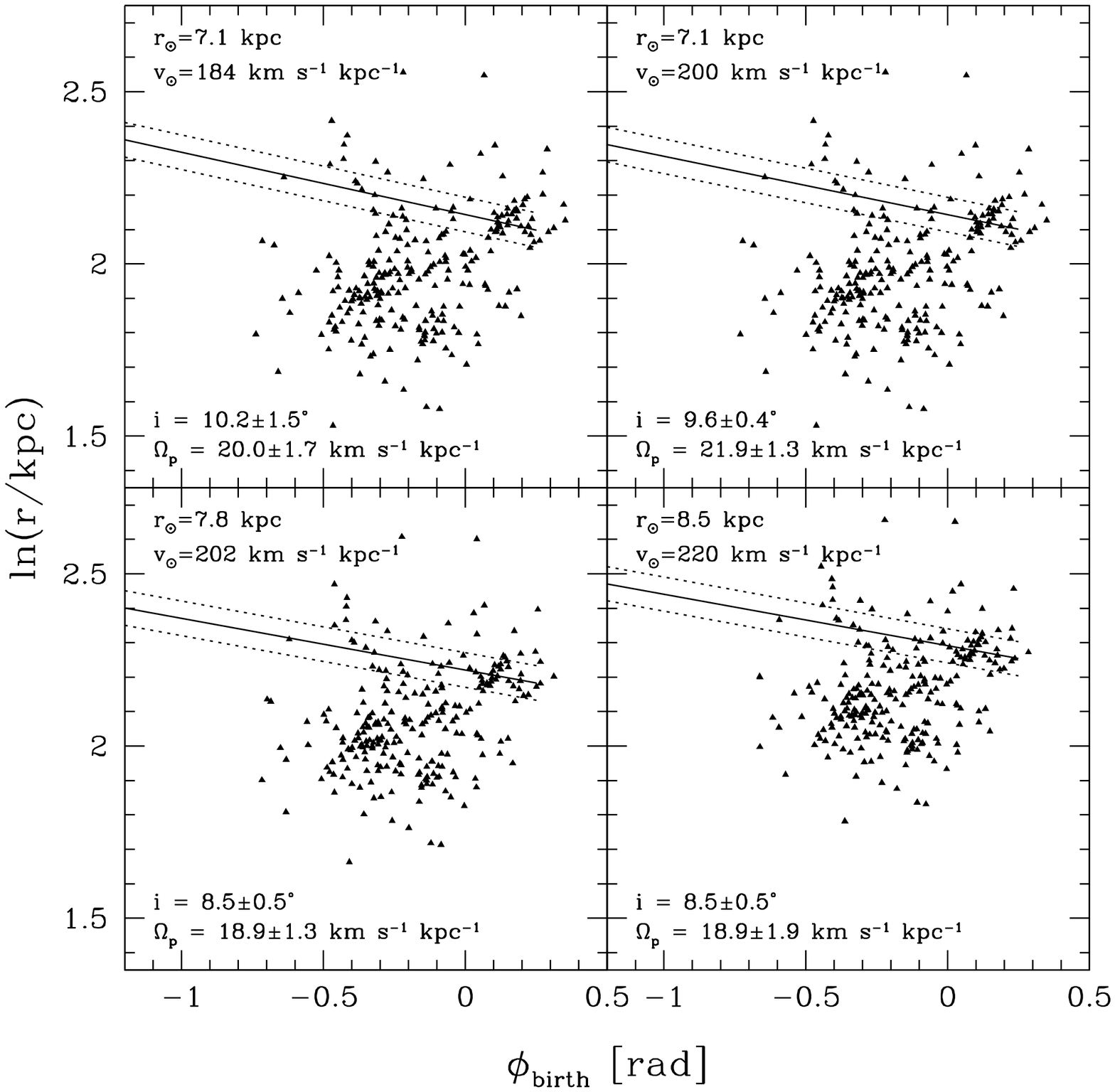}
\caption{A plot of $\ln r$ vs.\ the birth angle $\phi_\mathrm{birth}$ of clusters and the best fit for the Perseus arm, as described in fig.~\ref{fig:lower_firstarm}.}
\label{fig:upper}
\end{figure*}
}
\figurefive

\section{ The Dynamics of the Perseus Arm}\label{p}
When compared with the Carina arm, the Perseus arm includes less data points. Consequently the results we obtain are not as conclusive as we could hope for, though they are still statistically significant. The values found are summarize in table~\ref{tab:all} and fig.~\ref{fig:upper}.

The pattern speed obtained for the Perseus arm is somewhat higher than the lower Carina-arm pattern speed, though not totally consistent if one considers only the statistical error. This could arise from unaccounted systematic errors, as will be elaborated in the discussion. Moreover, if we repeat  the analysis while fixing the Perseus arm pitch angle with the somewhat higher value obtained for the slower Carina arm, then the best fit obtained for the Perseus arm has a $\sim \pi/2$ phase shift relative to the slower Carina arm. This strongly suggests that the two arms are indeed related to each other as part of an $m=4$ set.  If we repeat the analysis while fixing $\Omega_P$ for the Perseus arm with the value of the slower Carina arm, we again obtain a fit which has a $\sim \pi/2$ shift relative to the slower Carina arm.

\section{ The Dynamics of the Orion Arm}\label{orion}
The Sun is located in the vicinity of the Orion arm. The Orion arm itself is often regarded as an armlet, since it does not appear to be part of the Carina and the Perseus arms \citep{Georgelin}. Nevertheless, it is a site of star formation and has the features of a density wave. Irrespective of whether or not it is a full fledged spiral arm, or just a ``small'' pertrubation, we can estimate the pattern speed associated with it. 

Looking at fig.~\ref{fig:h_orion}, we can identify a high pattern speed for the entire range of the assumed  solar galactocentric radius and angular velocity. We should point out that for the $8.5$~kpc and the $7.8$~kpc configurations, we find two different  patterns, with very similar pattern speeds but somewhat different pitch angles. One may argue   that this indicates multiplicity  is the spiral structure, but since the two pattern speeds resemble in value, it is not a strong argument. This is because we only have relatively few cluster points to represent the Orion arm, such that the very few clusters at large radii can signficantly offset the pitch angle determination. Moreover, for a large assumed pitch angle for the Orion arm, the Orion and Perseus arms coincide, in which case Perseus clusters can accidentally be included in the pitch angle determination. Thus, we can only be certain of the existence of one set. The fit having the lower pitch angle (and lower pattern speed) is more favorable, while the second solution could be an artifact, because of the aforementioned coincidence.  

Irrespectively, we summarize our results for both fits of this arm, but the multiplicity should be taken with a grain of salt. An example fit to the Orion arm is portrayed in fig.~\ref{fig:o_78}, while the rest of the data is summarized in table \ref{tab:all}.

\def\figuresix{
\begin{figure}
\centering
\plotone{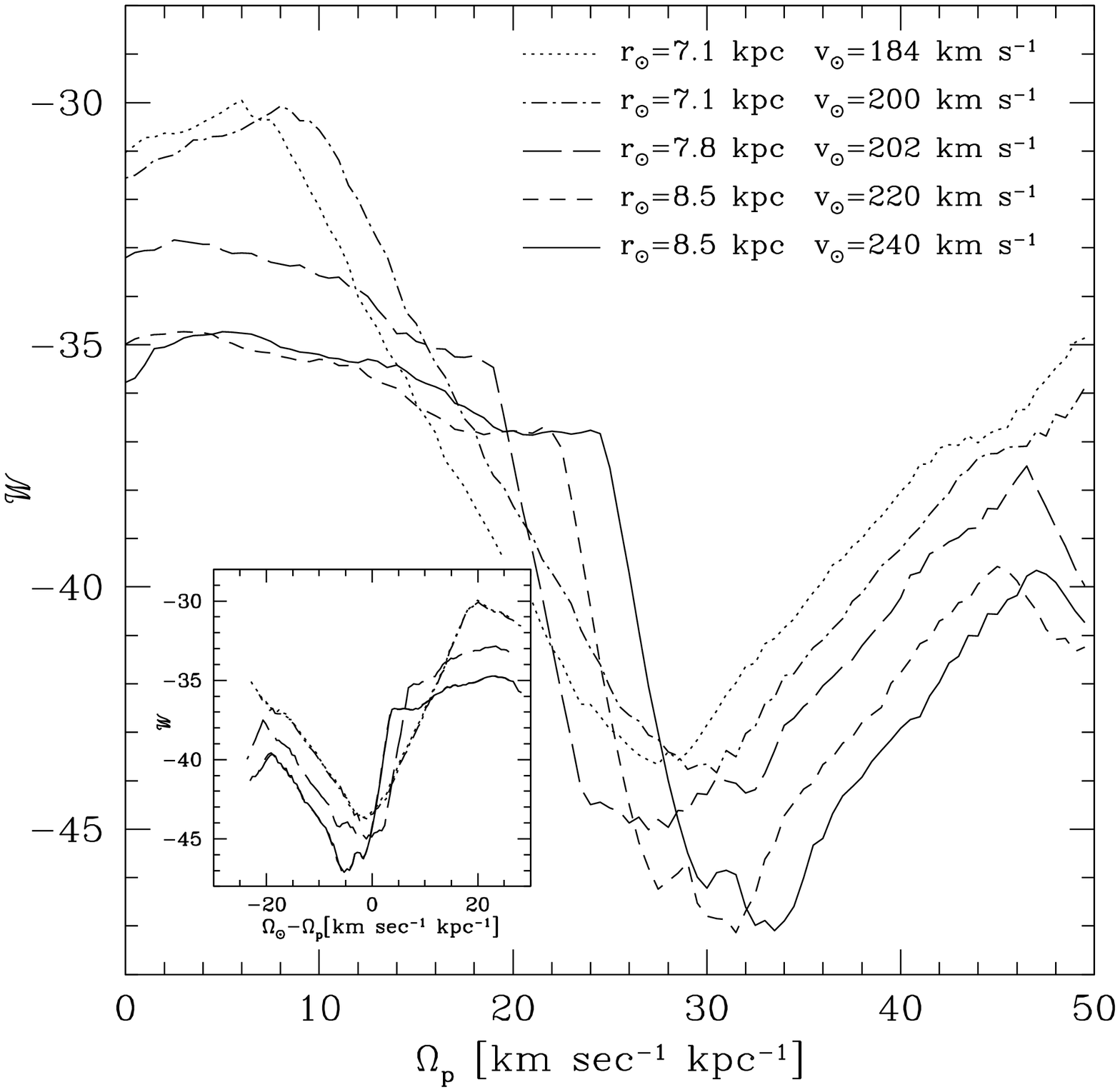}
\caption{\footnotesize The weight function ${\cal W}$ vs.\ the pattern speed $\Omega_P$ for the  Orion arm. The five different graphs represent the different assumed location and velocity of the sun with respect to the center of the galaxy. Either one or two nearby minima exist, for $\Omega_P \sim \Omega_\odot$. Namely, the solar system is near the co-rotation radius of the Orion arm. In the inset we plot ${\cal W}$ vs.\ $\Omega_\odot-\Omega_P$. Evidently, the method is more accurate in determining the relative pattern speed of the arms. Similarly to the Carina arm, we again find that configurations with $r_\odot \gtrsim 8~kpc$ give consistently better fits.}
\label{fig:h_orion}
\end{figure}
}
\figuresix

\def\figureseven{
\begin{figure}
\vskip -3.5cm
\centering
\epsfig{file=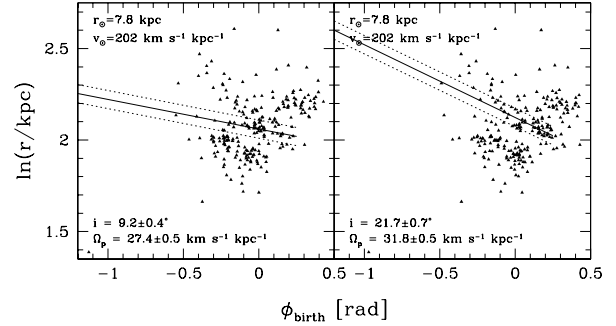,width=3.2in}
\caption{\footnotesize $\ln r$ vs.\ the birth angle $\phi$ with the two best fits for the Orion arm, while considering one configuration for the galaxy (similar to fig.~\ref{fig:lower_firstarm}).}
\label{fig:o_78}
\end{figure}
}
 \figureseven

\def\figureeight{
\begin{figure}
\vskip -0.5cm
\centering
 \plotone{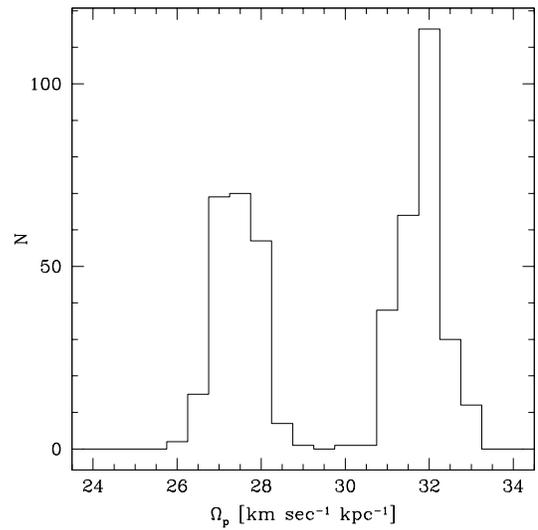}
\caption{\footnotesize A histogram
showing $\Omega_P$ of the best fits for the Orion arm, revealing the notable bi-modality.}
\label{fig:o_78}
\end{figure}
}
 \figureeight

\section{ Discussion }

\def\tableone{{
\begin{table*}
\caption{\footnotesize Spiral pattern speeds for given $r_\odot$ and $v_\odot$}
\centering
\begin{tabular}{  c    c   c   c   c   c }

\hline \hline 
$r_\odot$  [\small{kpc}]  & 7.1   & 7.1  & 8.5  & 8.5 & 7.8  \\
$v_\odot$ [\small{km sec$^{1}$}] & 184 & 200 & 220 & 240 & 202 \\ 
\hline \hline 
\multicolumn{4}{l}{ \it The First Carina Arm} \\
\hline 

$\Omega_P$ [\omunits] & $ 15.8\pm 0.9 $ & $ 17.9 \pm 0.7$ & $ 15.6\pm 0.9$ & $ 16.9 \pm 1.8$ & $ 15.3 \pm 1.6$ \\
\hline
$i$& $16.6 \pm 1.2^{\circ}$ & $16.2 \pm 1.3^{\circ}$ & $15.6 \pm 0.4^{\circ}$ & $14.9 \pm 1.9^{\circ}$ &  $15.7 \pm 1.5^{\circ}$ \\

\hline \hline
\multicolumn{4}{l}{ \it The Second Carina Arm} \\
\hline 
$\Omega_P$ [\omunits] & $ 29.7\pm 0.8 $ & $ 31.8 \pm 0.6$ & $ 27.6\pm 0.3$  & $ 30 \pm 0.5$ & $ 28.4 \pm 0.6 $ \\
\hline
$i$& $20.5 \pm 1.9^{\circ}$ & $21.1 \pm 1.7^{\circ}$ & $21.4 \pm 0.9^{\circ}$ & $21.5 \pm 1^{\circ}$ &  $22.4 \pm 1.1^{\circ}$ \\
\hline

\hline \hline
\multicolumn{4}{l}{ \it The Perseus Arm} \\
\hline 
$\Omega_P$  [\omunits]  & $ 20.0\pm 1.7 $ & $ 21.9 \pm 1.3$ & $ 18.9\pm 1.9$  & $ 21.0 \pm 1.8$ & $ 18.9 \pm 1.3 $ \\
\hline
$i$& $10.2 \pm 1.5^{\circ}$ & $9.4 \pm 0.4^{\circ}$ & $8.5 \pm 0.5^{\circ}$ & $8.5 \pm 0.5^{\circ}$ &  $8.5 \pm 0.5^{\circ}$ \\

\hline \hline
\multicolumn{4}{l}{ \it The Orion Arm} \\
\hline 

$\Omega_P$  [\omunits]  & $ 27.9\pm 0.5$ & $ 30 \pm 0.5 $ & $ 27.7\pm 0.5 $ & $ 30.2 \pm 1.6$ & $ 27.4 \pm 0.5$ \\
 & & &$31.5\pm0.5$ &$ 33.4 \pm 0.5$&$31.8\pm 0.5$\\
\hline
$i$& $10.6 \pm 0.6^{\circ}$ & $10.5 \pm 0.6^{\circ}$ & $8.5 \pm 0.7^{\circ}$ & $8.5 \pm 0.5^{\circ}$ &  $9.2 \pm 0.4^{\circ}$ \\
 & & & $21.3\pm0.9$& $21.3 \pm 0.9^{\circ}$ & $21.7 \pm 0.7^{\circ}$\\
\hline
\centering
\end{tabular}
\label{tab:all}
\end{table*}
}}

\tableone

In our analysis, we found two sets of spiral arms coexisting in the vicinity of what appears to be the single ``Carina arm". In the ``velocity spectrum" diagram, the lower $\Omega_P$ fit is more prominent, with lower values of the weight function ${\cal W}$ (see figure \ref{fig:h_lower}). It corresponds to a pattern speed of $\Omega_{P,\mr{Carina},1}=16.5^{+1.2}_{-1.4sys}\pm1.1_{stat}~\omunit,$ and a pitch angle of $i_{\mr{Carina},1}=15.8^{+0.8}_{-0.9sys}\pm1.2_{stat}^{\circ}$. The systematic error only includes the uncertainty arising from the inaccurate determination of the solar galactocentric radius and velocity. Other systematic errors could certainly exist, though hard or impossible to properly account. For example, the clusters we use are biased towards lower distances, implying that a higher weight is given to clusters on one side of the arm. In the case of the Carina arm, it is the older clusters which are favored. In the Perseus arm, it is the opposite. A second systematic error arises from our implicit assumption in eq.~\ref{birth} that the amplitude of the density wave is negligible. Alleviation of this assumption is done in a paper in progress. 

Since the measurement is essentially based on the speed of the star clusters {\em relative} to the orbital frequency at our galactic radius, a more accurate result is obtained if we fit for $\Delta\Omega=\Omega_\odot - \Omega_P$, such that the systematic error in $\Omega_\odot$ is less important. We find:
\begin{equation}
\Omega_\odot - \Omega_{P,\mr{Carina},1}= 10.6^{+0.7}_{-0.5sys} \pm 1.1_{stat}~\omunit.
\end{equation}
For comparison, the relative pattern speed $\Delta \Omega$ can be measured using the periodic variations in the cosmic ray flux history as recorded in Iron meteorites. This gives $\Delta \Omega =  10.5\pm1.5_{sys} \pm 0.8_{stat}~\omunit$ \citep{ice}. Alternatively, it can be compared to the periodicity obtained through the climatic effects of the variable cosmic ray flux. This geological determination results with  $\Delta \Omega =  10.4\pm1.5_{sys} \pm 0.35_{stat}~\omunit$ \citep{ice}. Thus, the lower  pattern we find for the Carina arm is in   agreement, within the measurement error,  with both a direct and indirect determination of the cosmic ray flux variability. Note that the {\em very} good agreement between the results obtained here and the meteoritic or geological determination should be considered a coincidence. The agreement is less impressive if one consolidates the results of the slower Carina arm with the Perseus arm, as discussed below. 

The $\Omega_{P,\mr{Carina},1}$ pattern speed places the sun in the vicinity of the inner (1:4) Lindblad resonance (see fig.~\ref{fig:resonances}).  This is consistent with the analysis of \cite{quillen}, who modeled stars in the solar vicinity and found that the observed dynamical signatures of chaos can be explained if the solar system is in the vicinity of an inner Lindblad resonance (either the 1:4 resonance for a 4 armed set or the 1:2 resonance for a 2-armed set). 
 
With reasonable certainty, we can also state that a weaker arm, possibly part of a larger set, is overlapping the previously described arm at the same radial range, but with a higher pattern speed. 
The second pattern speed we obtained for the Carina arm is $\Omega_{P,\mr{Carina},2}=29.8^{+0.6}_{-2.2 sys}\pm2.1_{stat}~\omunit$, with a pitch angle $i_{\mr{Carina},2}=21.1^{+1}_{-0.9sys}\pm1.3_{stat}^{\circ}.$
The statistical error here is smaller than for the faster arm, with the reason being the smaller dispersion associated with the smaller relative speed.  The relative pattern itself is 
\begin{equation}
\Omega_\odot - \Omega_{P,\mr{Carina},2}= -2.7 \pm 1_{sys} \pm 0.6_{stat}~\omunit.
\end{equation}
Again, we find that the relative pattern speed is better constrained than the absolute value.

The value of $\Omega_{P,\mr{Carina},2}$ places the observed outer extent of the Milky way's spiral structure  in the vicinity of the outer (1:2) Lindblad resonance. This also implies that we are just outside co-rotation for this spiral set.

Our results are in agreement with the two ranges of values often found in the literature, of $\sim 16-20~\omunit$ and $\sim 30~\omunit $.
For instance, \cite{ivanov} found $\Omega_P=16-20~\omunit$, using cluster age gradients. His result agrees with our first set, as well as with \cite{nelson}, \cite{Efremov}, and others. The second number is in nice agreement with previous determinations as well. For example, \cite{fre} found $\Omega_P=30\pm7~\omunit$, while \cite{corotation1} placed the solar system at $\Delta R \gtrsim 0.1$~kpc, relative to the co-rotation radius.

From linear density wave theory, spiral arms are expected to exist only between the Lindblad resonances depending on the particular pattern speed. Observationally, a clear 4-armed spiral set appears to exist at least from our galactocentric radius $r_\odot$, out to about twice this radius, $r_{out} \approx 2 r_\odot$ \citep{Blitz}.

When combining the above theoretical argument with observations, the outer extent of the faster set should end significantly inwards of the actual observed outer radius $r_{out}$, if the fast arm is part of a 4-armed set. On the other hand, it could extend out to $r_{out}$, if the arm is part of a 2-armed set (see fig.~\ref{fig:resonances}). In other words, the 4-arms observed out to $r_{out}$ cannot all belong to the faster set. Thus, one possibility is that both the slower and faster sets are 4 armed, in which case the slower set extends out to $r_{out}$ while the faster set necessarily  ends significantly inwards of $r_{out}$. A second possibility is that the slower set is still 4-armed but the faster set is 2-armed. Here, the observed 4 arms are still part of the slower set, however, superimposed on them, we have a faster 2-armed set that could extend as far out as $r_{out}$, but could also end further in. 

The inner extent of the observed spiral arms can be used to constrain the patten speeds from below. Inspection of fig.~\ref{fig:resonances} reveals that the inner extent of the slower set should be roughly at our galactic radius. In fact, the nominal lower limit is just outside the solar circle. Since we know the sun does pass through the arms of this set \citep[e.g., from meteoritic or geological data][]{ice}, a very small discrepancy of less than a few 100 pc exists. This should no be a source of major concern for several reasons. 

First, our knowledge of $\Omega_P$ which determines the inner Lindblad radius suffers from both systematic and statistical errors. Increasing ${\Omega_{P,1}}$ by its statistical error, for example, shifts the inner extent by more than 0.5 kpc.

Other interesting points to consider is that the sun performs epicyclic motion around a center located roughly $\sim 0.5$~kpc outside the current solar circle \citep[e.g.][]{matese}, allowing the sun to pass through the spiral arms even if it is currently just at their inner edge.

Near the resonance, the response of both the gas and stars to the spiral wave perturbation is extreme. This is expected to give rise to interesting nonlinear effects.  Moreover, in several numerical simulations and observation, spiral arms have been observed to exceed the inner, outer or both Lindblad resonance. For example, \cite{model1} have $H\alpha$ de-projected images of the galaxy NGC 157, demonstrating the existence of spiral arms beyond the predicted  outer  Lindblad resonance. \cite{armsC1} and \cite{armsC2} reported the existence of spiral patterns which exist inside the inner Lindblad resonance and connected to the outer pattern. This prompted \cite{shlosman} to analyze the formation of spiral patterns inside the Lindblad resonance. 
 
 \cite{sellwood} reached similar conclusions about spiral arms beyond the Lindblad resonances, but based his analysis on numerical simulations.  Furthermore, \cite{dwave1, dwave2} studied the propagation of waves in spiral galaxies. They argued that short density waves are not restricted to any region (whereas long density waves are restricted to exist between the Linndblad resonances). This is especially the case, if the origin is a rotating bar. 

To summarize the ``Carina arm", we can conclude that it is a superposition of two arms with two different pattern speeds. The slower arm is definitely 4-armed in structure and dominates the outer parts of the galaxy. On the other hand, nothing definitive can be concluded on the number of arms in the faster set, which dominates the galaxy within the solar circle.

 We can also conclude, more generally, that there is a real physical reason for the previous discrepancies between the different pattern speed determination---any pattern speed analysis which does not allow for the possibility of more than one set should be considered cautiously.

For the Perseus arm clusters, we find only one clear pattern speed having $\Omega_{P,\mr{Perseus}}=20.0 ^{+1.7}_{-1.2sys} \pm1.6_{stat}$\ $\omunit,$ with  a pitch angle of $i_{\mr{Perseus}}=9.0^{+1.8}_{-0.5sys}\pm0.7_{stat}^{\circ} .$ This result is consistent, for example, with \cite{omega20}, though a more careful comparison reveals some small discrepancy as described above. This could arise from yet unaccounted systematic errors. 

If we consider that the observed clusters in the Perseus arm span the range between $\gtrsim r_\odot$ to $\sim 1.5 r_\odot$, and that there is no sign for the $\Omega_{P,2}$ pattern  in the data, we can further extend our conclusion on the nature of the spiral arms, which we obtained using the Carina arm. Specifically, if the faster pattern is 4-armed, its outer extent  should be within corotation (and not, for example, the outer 4:1 Lindblad resonance), since corotation with $\Omega_{P,2}$ takes place in the vicinity of $r_\odot$. If the faster set is a 2-armed pattern, we do not expect an arm superimposed over the Perseus arm, and no conclusion can be reached regarding its outer extent.

Concentrating now on the Orion arm, we obtained that the best fitting arm has a pattern speed of $\Omega_{P,\mr{Orion},1}=28.9 ^{+1.3}_{-1.2sys}\pm0.8_{stat}~\omunit,$ with  a pitch angle of $i_{\mr{Orion},1}=9.5^{+0.9}_{-1.1sys}\pm0.6_{stat}^{\circ}.$ Again, the systematic error arises from our limited knowledge of the solar galactic radius and velocity. If we look at $\Delta \Omega$ instead, then the systematic error is notably smaller. We find:
$\Delta \Omega_{\mr{Orion},1} =1.8^{+0.2}_{-0.3sys}\pm0.7_{stat}~\omunit.$ 

However, under several assumption about the solar location and velocity, we find a second possibility, consistent with $\Delta \Omega_{\mr{Orion},1} =5.6^{+0.3}_{-0.4sys}\pm0.5_{stat} ~\omunit,$ and  a pitch angle of $i_{\mr{Orion},1}=21.4 ^{+0.3}_{-0.1sys}\pm0.8_{stat}^{\circ}.$

As previously explained, it is not clear whether the two solutions indeed imply that two pattern speeds exist. One possibility which seems reasonable, is that the second $\Delta \Omega$, with its large pitch angle, is obtained as an artificial fit where an ``arm" extends from the actual Orion arm clusters to the Perseus arm clusters, and is therefore an artifact.

Interestingly, the Orion and faster Carina sets have similar pattern speeds. It may simply be a coincidence, since there is no evidence indicating that they are part of the same set, however, there could be some kind of dynamic coupling. This is an interesting question to address in future research.



\end{document}